\documentstyle[prl,aps,twocolumn,epsfig]{revtex}
\begin{document}

\twocolumn[\hsize\textwidth\columnwidth\hsize\csname @twocolumnfalse\endcsname

\title{Pressure-Induced Intermediate-to-Low Spin State Transition in LaCoO$_3$}
\author{T. Vogt$^{1,*}$, J.A. Hriljac$^2$, N.C. Hyatt$^3$, P. Woodward$^4$}
\address{$^1$Physics Department,Brookhaven National Laboratory,Upton, NY 11973-5000,USA}
\address{$^2$School of Chemistry, The University of Birmingham, Edgbaston, Birmingham, B15 2TT, United Kingdom}
\address{$^3$Department of Engineering Materials, The University of Sheffield,Sheffield, S1 3JD, UK}
\address{$^4$Department of Chemsitry, Ohio State University, Columbus, Ohio 43210, USA}
\date{\today}
\maketitle

\begin{abstract}

Synchrotron X-ray powder diffraction experiments reveal that the transition
from a magnetic intermediate spin (IS) state t$^5$$_{2g}$e$^1$$_g$ to a
nonmagnetic low-spin (LS) ground state t$^6$$_{2g}$ in LaCoO$_3$ normally
observed when cooling, manifests itself under pressure by an anomalously low bulk compressibility of
150(2) GPa and an initially very large Co-O bond compressibility of 4.8 x 10$^{-3}$ 
GPa$^{-1}$ which levels off near 4 GPa. The continuous depopulation
of the IS state is driven by an increased crystal field splitting resulting in an effective reduction
of the size of the Co$^{3+}$ cation.

\end{abstract}

\vskip1pc]

%\newpage
\narrowtext

% === main text ===
The magnetic and electronic properties of the paradigmatic charge-transfer
insulator LaCoO$_{3}$ which crystallizes in a rhombohedral distortion of the
cubic perovskite structure with the a$^{-}$a$^{-}$a$^{-}$ tilt system \cite
{Glazer} continue to be topical. Magnetic susceptibility measurements reveal
transitions at 100 and 500K. The 100K transition was first interpreted by
Goodenough\cite{Goodenough} as a spin state transition of 50\% of the Co$%
^{3+}$ ions from a nonmagnetic low-spin (LS, t$_{2g}^{6}$, S=0) ground state
to a high spin state (HS, t$_{2g}^{4}$e$_{g}^{2}$, S=2). The 500K transition
was assigned to an order-disorder transition, where the e$_{g}$ electrons
become itinerant and destroy the spin-state ordered superstructure the
latter, however, was never observed experimentally\cite{thornton1},\cite
{thornton2} and a dynamic disorder of HS and LS Co$^{3+}$ was subsequently
proposed\cite{senaris}. An alternative interpretation based on
photo-electron emission \cite{masuda} and X-ray absorption spectroscopy\cite
{abbate},\cite{saitoh} led to the postulation of an intermediate spin (IS, t$%
_{2g}^{5}$e$_{g}^{1}$, S=1) state and the two transitions were assigned to
thermally activated LS-to-IS and IS-to-HS state transitions\cite{asai}.
According to classical ligand field theory the IS state should always be
energetically above the LS or HS state. However, LDA+U calculations by
Korotin et al\cite{korotin} established that the IS state could be
energetically stabilized with respect to the HS state. Recent neutron powder
diffraction experiments by Radaelli and Cheong \cite{radaelli} show that the
thermal lattice expansion is best fitted by such a LS-to-IS-to-HS state
sequence without orbital degeneracy of the IS state. However, correction
terms to a simple activated behavior and a significantly reduced HS
effective moment are required, indicating that this model is, at present,
incomplete.

Pressure can alter the magnetism of transition-metal oxides by inducing Mott 
\cite{pasternak1} or spin state transitions\cite{pasternak2}. The first is a
consequence of the closure of the Mott-Hubbard or charge-transfer gap,
whereas the latter results from a breakdown of strong Hund's rule coupling
and occurs when the crystal field splitting dominates over the exchange
energy. A second order HS-to-LS transition under pressure leads to the
collapse of the magnetic state in wuestite (Fe$_{0.94}$O) as shown by
M\"{o}ssbauer spectroscopy\cite{pasternak1}. In RFeO$_{3}$ (R=La, Pr) a
pressure-induced collapse of the Mott-Hubbard state and a subsequent
continuous HS-to-LS transition was observed\cite{xu}. Takano et al observed
a first-order pressure-driven HS-to-LS transition in CaFeO$_{3}$ by in-situ
M\"{o}ssbauer spectroscopy and x-ray diffraction\cite{Takano}. In the
following we present experimental evidence for a continuous IS-to-LS state
transition occurring in LaCoO$_{3}$ under hydrostatic pressure at room
temperature.

The experimental setup and detailed procedure for the high-pressure
synchrotron x-ray powder diffraction experiments performed at beam line X7A
at the National Synchrotron Light Source at Brookhaven National Laboratory
are described elsewhere\cite{vogt}. Rietveld refinements were performed
using the program GSAS\cite{larson}. The results of the fits are summarized
in Table 1. The lattice parameters and the volume of the unit cell reveal no
discontinuity as a function of pressure (Figure 1). Intriguingly however,
the value of the bulk modulus, B$_{0}$, of 150(2) GPa obtained by fitting 
\cite{EOSfit} a second order Birch-Murnagham Equation of State\cite{birch}
with V$_{0}$=335.834(5)\AA $^{3}$ and B'=4 (by definition) to the pressure
dependence of the unit cell volume is significantly smaller than those
observed for comparable perovskites with the a$^{-}$a$^{-}$a$^{-}$ tilt
system such as LaAlO$_{3}$ (B$_{0}$=190(5) GPa)\cite{bouvier} and PrAlO$_{3}$
(B$_{0}$=205(8) GPa) \cite{kennedy}. The Cornellius-Schilling model \cite
{cornellius} predicts values of 180, 189 and 191 GPa for B$_{0}$ of LaCoO$%
_{3}$, LaAlO$_{3}$ and PrAlO$_{3}$ respectively. Our experimental value is
therefore highly indicative of an unusually low volume compressibility in
LaCoO$_{3}$.

The La$^{3+}$ cation in LaCoO$_{3}$ is at the center of a distorted
dodecahedron and coordinated by twelve oxygen atoms with 3 long, 6
intermediate and 3 short distances (Figure 2)\cite{woodward}.The three
closest oxygen atoms to La are coplanar and the La p$_{x}$- and p$_{y}$%
-orbitals point at this plane. The p$_{z}$ and d$_{z^{2}}$ orbitals are
orthogonal to this plane and have no overlap with the first coordination
sphere of oxygen. The majority of the La-O bonding stems from the three
short La-O bonds. As pointed out by Radaelli and Cheong\cite{radaelli} and
others, the long and short La-O bonds are useful gauges for non-thermal
lattice distortions. An increase in the long La-O distance accompanied by a
decrease of the short one, leading to a more distorted environment for La,
is indicative of an increase in the magnitude of the octahedral distortion.
The Co$^{3+}$ cation is at the center of a distorted octahedron surrounded
by six equivalent oxygen atoms.

The effect of pressure on the individual metal-oxygen bond distances clearly
reveals that there are two distinct regions with a change in slope near 4
GPa (Figures 3a and b). Initially the short La-O bonds increase, the long
ones decrease, and the six intermediate ones also decrease, but by a much
smaller amount. Under pressure, the Co-O bonds contract, as expected. When
analyzing the individual linear bond compressibilities ($\beta
_{L}=(-1/L)(\delta L/\delta P)_{T})$ up to 4 GPa, one observes that the
contraction of the intermediate La-O bond ( $\beta
_{La-O3}=3.6x10^{-3}GPa^{-1}$) matches the Co-O bond compressibility ($\beta
_{Co-O}=4.8x10^{-3}GPa^{-1}$ ). In contrast, the three long La-O bonds
contract very strongly ($\beta _{La-O1}=1.8$x10$^{-2}$ GPa$^{-1}$), while
the short ones expand ( $\beta _{La-O2}=-1.76x10^{-2}GPa^{-1}$) at a
comparable rate. This is a clear indication of significant changes in the
magnitude of the octahedral tilting as discussed in the following paragraph.

An exceedingly useful measure of the distortion of a perovskite-type
structure from the cubic aristotype ABO$_{3}$ is the tilt angle, $\phi ,$
which gives the degree of rotation around the threefold axis of the BO$_{6}$
octahedra. O Keefe et al\cite{okeefe} have shown that the variation of $\phi 
$ with pressure in orthorhombic perovskites depends on the strength of the
two bond compressibilities $\beta _{A-O}$ and $\beta _{B-O}$. If the A site
is more compressible, $\beta _{A-O}>\beta _{B-O}$, $\phi $ should increase
with pressure and the structure will distort away from cubic symmetry. If
the bond compressibilities are reversed in magnitude, the tilt angle should
decrease with pressure. These trends are indeed observed in the three
isostructural orthorhombic systems (Pbnm, b$^{-}$b$^{-}$a$^{+}$) where $\phi 
$ increases in MgSiO$_{3}$\cite{kudoh} ($\beta
_{Mg-O}=1.1x10^{-3}GPa^{-1}>\beta _{%
%TCIMACRO{\func{Si}}%
%BeginExpansion
\mathop{\rm Si}%
%EndExpansion
-O}=8.9x10^{-4}GPa^{-1})$ and decreases in LaMnO$_{3}$\cite{pinsard},\cite
{loa} ($\beta _{La-O}=2.8x10^{-3}GPa^{-1}<\beta _{Mn-O}=3.2x10^{-3}GPa^{-1})$
and YAlO$_{3}$\cite{ross1}($\beta _{Y-O}=2.0x10^{_{-3}}GPa^{-1}<\beta
A_{l-O}=2.5x10^{-3}GPa^{-1})$.In the related system ScAlO$_{3}$, the
compressibilities of $\beta _{Sc-O}$ and $\beta _{Al-O}$ are essentially
equal ($\simeq 1.5x10^{-3}GPa^{-1})$, and there is no significant change in $%
\phi $ with pressure\cite{ross2}. To our knowledge, the only previously
published detailed crystallographic data for a rhombohedral perovskite under
pressure is the case of PrAlO$_{3}$\cite{kennedy}, which is isostructural
with LaCoO$_{3}$ and also adopts the a$^{-}$a$^{-}$a$^{-}$ tilt system
(space group R-3c) up to ca. 7 GPa. Above this pressure PrAlO$_{3}$
undergoes a phase transition into the a$^{0}$b$^{-}$b$^{-}$ tilt system
(space group Imma). Similar to LaMnO$_{3}$ and YAlO$_{3}$, the A-O bond is
less compressible ($\beta _{\Pr -O}=2x10^{-3}GPa^{-1}<\beta
_{Al-O}=2.6x10^{-3}GPa^{-1})$ and the tilt angle decreases with pressure.
When considering the data to 4 GPa for LaCoO$_{3}$,once again the A site is
less compressible ($\beta _{La-O}=3.0x10^{-3}GPa^{-1}$) than the B-site ($%
\beta _{Co-O}=4.8x10^{-3}GPa^{-1})$ and the tilt angle decreases. Therefore
it appears that the argument forwarded by O'Keefe et al \cite{okeefe}
relating individual bond compressibilities to a distortion under pressure is
also valid for perovskites that adopt other than orthorhombic crystal
symmetry.

The variation of the tilt angle as well as the Co-O-Co angles for LaCoO$_{3}$
under pressure (Figure 4) clearly emphasize what has already been noted from
the pressure dependence of the bond distances, that, in contrast to the unit
cell parameters there is an obvious change in the response of the bond
distances and angles above ca. 4 GPa. Based on a comparison of the limited
number of experimentally determined individual bond compressibilities in
perovskites, it is clear that below 4 GPa, the Co-O bond is highly
compressible ($\beta _{Co-O}=4.8x10^{-3}GPa^{-1})$. In fact it is {\it the
most compressible B-O bond observed} to date (greater than Si$^{4+}$-O, Mn$%
^{3+}$-O and Al$^{3+}$-O). Above 4 GPa, the compressibility is considerably
lower ($\beta _{Co-O}=1.2x10^{-3}GPa^{-1})$, much more in line with the
expected value for a B-O bond in a perovskite. To place these observations
within our current framework of understanding one needs to take into account
the depopulation of the IS state under pressure. Asai et al\cite{asai2} have
shown experimentally that the energy gap between LS and IS states increases
under pressure. At pressures above 0.5 GPa this increase is found to be
quadratic indicating a very strong volume dependence of the IS state energy.
Therefore the IS state will be depopulated under pressure and as a result a
small but noticeable contraction of the Co-O bond is to be expected. This
contraction occurs as a result of the depopulation of Co e$_{g}$ orbitals,
which are Co-O $\sigma $-antibonding in character. This is the origin of the
well-known fact that LS Co$^{3+}$ has a smaller radius (0.685\AA ) than HS Co%
$^{3+}$ (0.75\AA )\cite{shannon}. If we estimate the radius of IS Co$^{3+}$
as the arithmetic mean of LS and HS values (0.717\AA ), it is gratifying to
note that the expected IS Co$^{3+}$-O distance (1.927\AA ) is close to the
value of 1.933(1)\AA\ observed at ambient pressure. The expected LS Co$^{3+}$%
-O distance (1.895\AA ) is very close to 1.896(1)\AA , the value observed at
3.98 GPa as well as the zero-pressure value of 1.905\AA\ extrapolated back
from the region between 4 and 8 GPa\cite{comment}. Obviously, the fairly
rapid and continuous contraction of the effective size of the cobalt
centered octahedra leads to a decrease in the magnitude of the octahedral
tilting distortion. Both the observed Co-O distances and the changes in the
octahedral tilting strongly suggest that at 4 GPa, the IS is significantly
depopulated. The dramatic change in Co-O bond compressibility as well as the
change from a decreasing to an increasing tilt angle that occurs above 4
GPa, suggests that the spin state transition for IS to LS is essentially
complete at 4 GPa. The fact that at higher pressures the LS Co$^{3+}$-O
bonds have a lower compressibility than the La-O bonds is responsible for
the gradual increase in octahedral tilting between 4 and 8 GPa.

In summary, we have shown that LaCoO$_{3}$ has an anomalously low bulk
compressibility (B$_{0}$=150(2) GPa). Furthermore, individual bond
compressibilities reveal a change in their pressure-dependence near 4 GPa
indicating a change in the compression mechanism. This pressure behavior is
caused by an 'unusually' compressible Co-O bond. We argue that the reason
for this is a pressure-induced continuous depopulation of the IS state over
the 0 - 4 GPa pressure range. This depopulation is driven by the increased
crystal field splitting between the t$_{2g}$ and e$_{g}$orbitals, which
results in a reduction of the effective size of Co$^{3+}$ as antibonding
electrons are removed from the e$_{g}$ orbitals. We encourage
first-principles calculations of the electronic structure taking into
account this spin state transition to compare with our experimental values.
This is to our knowledge the first observation of a continuous
pressure-induced IS-to-LS transition under pressure in a perovskite.
Furthermore, we have demonstrated that high-pressure diffraction provides an
excellent opportunity to probe the structural response of spin state
transitions.

We gratefully acknowledge discussions with Fabian Essler and John
Tranquada.and would like to thank J. Hu from the Geophysical Laboratory,
Carnegie Institution of Washington for being able to use the laser system at
beam line X17C for pressure determination. The work was supported by the
Division of Materials Science, US Department of Energy, under Contract NO.
DE\_AC02-98CH10886

* Corresponding author , Email address: tvogt@bnl.gov

\end{document}